\documentclass[showpacs,preprintnumbers]{revtex4}
\usepackage{graphicx}
\usepackage{dcolumn}
\usepackage{bm}
\usepackage{epsfig}
\usepackage{amsfonts}
\usepackage{graphicx}
\usepackage{amsmath}

\begin{document}

\title{Stability Analysis of Canonical Acoustic Black Hole: The Role of Non-Minimal Scalar Coupling}

\author{ \firstname{HengZhong} \surname{Fang} }
\email{fanghzh@upc.edu.cn}
\affiliation{Department of Physics, China University of Petroleum, Qingdao 266580, China }
\author{ \firstname{LiQiang} \surname{Miao} }
\email{1246930652@qq.com}
\affiliation{Department of Physics, China University of Petroleum, Qingdao 266580, China }
\author{ \firstname{ChangSen} \surname{Liu} }
\email{qdlcsen@163.com }
\affiliation{Department of Physics, China University of Petroleum, Qingdao 266580, China }
\author{ \firstname{MengXue} \surname{Zhang} }
\email{2409020207@s.upc.edu.cn}
\affiliation{Department of Physics, China University of Petroleum, Qingdao 266580, China }

\begin{abstract}
In this paper, we investigate the stability of black hole spacetimes under the perturbation of a non-minimally coupled scalar field, using the canonical acoustic black hole as a concrete representative background. The non-minimal coupling parameter \(\zeta\), inspired by the standard \(\xi R \phi^2\) term in curved spacetime quantum field theory, is introduced phenomenologically to probe the effect of the effective curvature. By employing the sixth-order WKB approximation, we compute the quasi-normal mode (QNM) frequencies for various multipole numbers \(l\) and overtones \(n\). Our analytical analysis identifies a tachyonic instability threshold at \(\zeta_{\text{crit}} = -2/3\) for the monopole perturbation. For \(\zeta \ge -2/3\), the effective potential is strictly positive and no tachyonic instability of this type is expected. For \(\zeta < -2/3\), a negative potential well emerges, which may give rise to a tachyonic instability, as confirmed by our time-domain simulations. We further compare the scalar field results with those of the electromagnetic and Dirac fields, which serve as minimally coupled baselines. The contrast suggests that the tachyonic instability is a feature arising from the non-minimal curvature coupling in the scalar sector within the parameter regime explored here. Although the quantitative threshold is background-dependent, the qualitative mechanism is expected to be generic to black hole backgrounds with non-minimal scalar coupling.
\end{abstract}

\maketitle

\section{Introduction}

General Relativity (GR) is accepted as the best classical description of gravitation. One of its most striking predictions is the existence of black holes (BHs), objects from which nothing, not even light, can escape once crossing the event horizon. The interest in BHs goes far beyond astrophysics: they serve as theoretical laboratories for understanding the fundamental interplay between quantum physics and gravitation. In the strong gravity regime near the horizon, various phenomena have been extensively analyzed, such as Hawking radiation\cite{Hawk,Gibb}, superradiance, and quasinormal modes (QNMs).

The study of the decay of small perturbations has a long history in general relativity. The stability of black holes was first investigated by Regge and Wheeler in their pioneering work\cite{Regg} on linear perturbations of the exterior Schwarzschild spacetime. When a perturbed system relaxes towards equilibrium, it does so through the emission of QNMs. These are solutions to the linearized equation of motion that obey purely outgoing (transmissive) boundary conditions at both spatial infinity and the event horizon. The frequencies of QNMs are complex, with the real part determining the oscillation frequency and the imaginary part dictating the damping time. Importantly, they depend only on the intrinsic parameters of the black hole (mass, charge, angular momentum), being independent of the initial perturbation that excited them. The detailed study of QNMs was initiated by Vishveshwara\cite{Vish}, while Press\cite{Pres1} coined the term "quasi-normal frequencies". 

In the study of general relativistic black holes, QNMs are of paramount importance. First, they provide a direct means to infer black hole parameters, which dates back to the 1970s\cite{Pres2,Sath,Eche,Thor,Card3}. With the recent advancements in gravitational wave astronomy\cite{Abbo1,Abbo2,Abbo3,Coll}, particularly the detection of ringdown signals from binary black hole mergers, this long-held goal has become a reality\cite{Yune}. Second, QNMs test the stability of the system, an imaginary frequency with the wrong sign would signify an exponentially growing mode rather than damping. Recently, investigations into the spectral stability of black holes have demonstrated the susceptibility of the QNM spectrum to small, local perturbations in the scattering potential\cite{Noll,Agui,Noll2,Yang}. Furthermore, the influence of various higher-order effects—such as environmental factors\cite{Cann,Leon}, linear metric perturbations\cite{Leav1,Leav2,Ande,Berti,Zhang,Hugh,Lim,Oshi,Lim2}, and nonlinearities\cite{Cheu}—on the ringdown phase has been extensively explored.

Analogue gravity provides a powerful experimental and theoretical platform to probe classical and quantum phenomena in curved spacetimes. It was Unruh who first established the analogy between the propagation of sound waves in a moving fluid and the kinematics of waves in a classical gravitational field\cite{Unru1}. In this framework, the background fluid flow acts as an effective black hole metric, leading to the concept of a "dumb hole" with a sound horizon where the flow velocity equals the speed of sound. These acoustic analogues exhibit remarkable structural similarities to astrophysical black holes, giving rise to phenomena such as phonon radiation (the analogue of Hawking radiation)\cite{Jaco,Unru2,Matt1,Fisc} and superradiance\cite{Rous,Wein,Euve}. The mathematical foundation of this analogy was rigorously formulated by Visser\cite{Matt1,Matt2}, who mapped the propagation of sound waves in a non-relativistic, irrotational, and barotropic fluid to a minimally coupled massless scalar field in a Lorentzian geometry. Over the past two decades, the phenomenology of specific models, such as the draining bathtub, has been extensively studied. Works on QNMs\cite{Card2,Dola}, absorption processes\cite{Oliv}, and superradiance\cite{Basa,Rich} have shown that these vortex geometries share many dynamical properties with Kerr black holes, which are known to be stable against linear bosonic perturbations\cite{Whit,Bert,Teuk}.

Since QNMs appear naturally in general relativistic black holes, they should also be present in their acoustic analogues. While previous studies have extensively analyzed the QNMs of minimally coupled fields in acoustic black hole\cite{Maty,Camp}s, the investigation of non-minimally coupled scalar fields remains limited. In classical field theory and cosmology, non-minimal coupling terms (such as \(\xi R \phi^2\)) are well-motivated, arising naturally from quantum corrections in curved spacetime or from conformal invariance requirements. Within the analogue gravity framework, a pioneering study by Chen and Jing \cite{Chen} calculated the QNMs of a non-minimally coupled scalar field in the canonical acoustic black hole using a third-order WKB approximation. However, their analysis was restricted to the regime of positive coupling parameters. The behavior in the negative coupling regime, particularly the onset of potential instabilities, has not yet been systematically explored. In this paper, we fill this gap by extending the analysis to arbitrary coupling values, where we analytically identify a tachyonic instability threshold and numerically confirm its dynamical consequences.

It is worth emphasizing that the analysis strategy adopted here is not restricted to the canonical acoustic black hole. The central ingredient---examining the sign structure of the near-horizon effective potential for a non-minimally coupled scalar field---is applicable to a broad class of black hole backgrounds in which a scalar field couples to the effective curvature. The canonical acoustic black hole is used in this work as a concrete representative background that allows the mechanism to be displayed explicitly and checked numerically. For other backgrounds, only the explicit forms of the metric function $f(r)$ and the Ricci scalar $R(r)$ change, while the qualitative logic of the sign analysis remains the same.

In this paper, we investigate the QNMs of a canonical acoustic black hole under the perturbations of a non-minimally coupled scalar field, as well as electromagnetic and Dirac fields. The non-minimal coupling parameter \(\zeta\) is introduced phenomenologically to probe the effective curvature effects. We analytically derive a tachyonic instability threshold and numerically compute the QNM frequencies using the sixth-order WKB approximation. The electromagnetic and Dirac fields are treated as minimally coupled baselines to isolate the coupling effects. It is worth noting that within the analogue gravity framework, the scalar field represents the acoustic perturbation of the fluid, while the electromagnetic and Dirac fields are introduced as test fields to probe the effective geometry. The non-minimal coupling \(\zeta\) is treated as a phenomenological parameter to investigate curvature-like effects on the acoustic waves.

The paper is organized as follows: In Sec. II, the metric of the canonical acoustic black hole is introduced. In Sec. III, we investigate the effective potential and stability under the perturbation of the non-minimally coupled scalar field. In Sec. IV and V, the stability analyses for the electromagnetic and Dirac fields are presented as comparative baselines. Finally, the conclusion is given in Sec. VI.

\section{The acoustic metric of a convergent flow}

From geometrical acoustics (the unitarity of the sound cone) or physical acoustics (the continuity equation and Euler's equation), the acoustic metric is either read off by inspection\cite{Matt3,Matt4}
\begin{equation}
g_{\mu \nu }=\frac{\rho }{c_{s}}\left[ \begin{array}{cc}
-\left( c_{s}^{2}-\upsilon ^{2}\right) & \overset{\rightharpoonup }{\upsilon}^{T} \\ 
\overset{\rightharpoonup }{\upsilon } & I_{3\times 3}
\end{array}\right]  \label{e1}
\end{equation}
where $\rho$ is the background density, $c_{s}$ is the speed of sound, $\overset{\rightharpoonup }{\upsilon}$ is the velocity of the fluid, and $I_{3\times 3}$ is the unit matrix. Equivalently, the acoustic line element can be expressed as
\begin{equation}
ds^{2}\equiv \ g_{\mu \nu }dx^{\mu }dx^{\nu }=\frac{\rho }{c_{s}}\left[-c_{s}^{2}d\tau ^{2}+\left( dx^{i}-\upsilon ^{i}d\tau \right) \delta_{ij}\left( dx^{j}-\upsilon ^{j}d\tau \right) \right]  \label{e2}
\end{equation}

For spherically symmetric, incompressible, convergent flow, $\rho$ is a position-independent constant throughout the flow. The equation of continuity then implies that the fluid speed (only the radial component nonzero) $\upsilon \propto 1/r^{2}$. For convenience, we choose
\begin{equation}
\upsilon =c_{s}\frac{r_{0}^{2}}{r^{2}},  \label{e3}
\end{equation}
where $r_{0}$ is a normalization constant corresponding to the position of the horizon of the convergent acoustic black hole where $\upsilon =c_{s}$.

Substituting into Eq. (\ref{e2}) and performing a coordinate transformation
\begin{equation}
dt=d\tau \pm \frac{\frac{r_{0}^{2}}{r^{2}}}{c_{s}\left[ 1-\left( \frac{r_{0}^{4}}{r^{4}}\right) \right] }dr,  \label{e4}
\end{equation}
we get the line element of the spherical convergent acoustic black hole as
\begin{equation}
ds^{2}=-c_{s}^{2}\left[ 1-\left( \frac{r_{0}^{4}}{r^{4}}\right) \right]dt^{2}+\frac{dr^{2}}{1-\left( \frac{r_{0}^{4}}{r^{4}}\right) }+r^{2}d\theta^{2}+{\sin^{2}}\theta d\phi ^{2}.  \label{e5}
\end{equation}
For simplicity, rewrite the metric above as
\begin{equation}
ds^{2}=-f\left( r\right) dt^{2}+\frac{dr^{2}}{f\left( r\right)}+r^{2}d\theta ^{2}+{\sin^{2}}\theta d\phi ^{2},  \label{ee6}
\end{equation}
where $f\left( r\right) =1-(r_{0}^{4}/r^{4}\ )$ and here we take the natural unit that $c_{s}=1$.

Eq. (\ref{e5}) describes a convergent spherical acoustic black hole metric with the horizon at $r=r_{0}$, called canonical acoustic black hole\cite{Matt1,Bert2}. It is worth noting that this canonical acoustic metric has a straightforward experimental realization: its time-dependent version exactly describes the acoustic metric around a spherically symmetric bubble with an oscillating radius\cite{Hoch}. Hawking radiation and information conservation for this convergent acoustic black hole were investigated by Fang and Zhou\cite{Fang} using a quasi-classical method issued by Parikh and Wilczek\cite{Par}.

\section{Scalar field perturbation to the acoustic black hole}

The equation of motion governing an acoustic perturbation in the velocity potential of an irrotational flow of a barotropic and inviscid fluid is the same as the Klein-Gordon equation for a massless scalar field propagating in a Lorentzian geometry. In this section, we consider a massless scalar field non-minimally coupled to the background geometry. Its dynamics is governed by the modified Klein-Gordon equation,
\begin{equation}
\frac{1}{\sqrt{-g}}\partial _{\mu }\left( \sqrt{-g}g^{\mu \nu }\partial _{\nu }\right) \mathrm{\Psi }-\zeta R\mathrm{\Psi }=0,  \label{e6}
\end{equation}
where \(R\) is the Ricci scalar of the acoustic metric. We adopt the convention \((\Box-\zeta R)\Psi=0\), in which the conformally coupled case in four dimensions corresponds to \(\zeta=1/6\). (Equivalently, in the alternative convention \((\Box+\xi R)\Psi=0\), the conformal value is \(\xi=-1/6\).) The non-minimal coupling parameter \(\zeta\) is introduced phenomenologically to probe effective curvature effects, similar to models used in cosmological contexts\cite{Abre,Mann,Arty}. Two values of \(\zeta\) are of particular interest: \(\zeta=0\) (minimally coupled) and \(\zeta=1/6\) (conformally coupled).

Here, we will start by developing Eq. (\ref{e6}) considering the background metric given in Eq. (\ref{e5}). Due to the spherical symmetry of the metric, we introduce the scalar field through standard ansatz in the following form:
\begin{equation}
\Psi =\frac{e^{-i\omega t}\varphi \left( r\right) }{r}Y_{lm}\left( \theta,\phi \right) ,  \label{e7}
\end{equation}
where \(Y_{lm}\left( \theta ,\phi \right) \) denotes the spherical harmonics functions, with \(l\) the angular quantum number and \(m\) the magnetic quantum number, respectively.

Performing a tortoise transformation
\begin{equation}
dr_{\ast }=\frac{dr}{f\left( r\right) },  \label{e8}
\end{equation}
which maps the acoustic horizon at \(r=r_{0}\) to \(r_{\ast }\rightarrow-\infty \) , and \(r\rightarrow +\infty \) to \(r_{\ast }\rightarrow +\infty \). Then together with Eq. (\ref{e5}) and Eq. (\ref{e6}), we got that the radial component of the coupled scalar field satisfies a Schr\"{o}dinger-like wave equation
\begin{equation}
\frac{d^{2}\varphi (r)}{dr_{\ast }^{2}}+\left( \omega ^{2}-V\left( r\right)\right) \varphi (r)=0,  \label{e9}
\end{equation}
and, the effective potential
\begin{equation}
V(r)=\left( 1-\frac{r_{0}^{4}}{r^{4}}\right) \left[ \frac{l(l+1)}{r^{2}}+\frac{4r_{0}^{4}}{r^{6}}+\zeta R\right] .  \label{e10}
\end{equation}

For the metric of the canonical acoustic black hole in Eq. (\ref{e5}), the Ricci scalar is
\begin{equation}
R=g^{\mu \nu }R_{\mu \nu }=\frac{6r_{0}^{4}}{r^{6}}.  \label{e12}
\end{equation}
So we have
\begin{equation}
V\left ( r \right ) = \left ( 1-\frac{r_{0}^{4} }{r^{4} } \right ) \left [ \frac{l\left ( l+1 \right ) }{r^{2} } +\frac{4r_{0}^{4} }{r^{6} }+\frac{6\zeta r_{0}^{4} }{r^{6} } \right ] . \label{e13}
\end{equation}

\subsection{Effective potential and tachyonic instability}

From Eq. (\ref{e13}), the effective potential depends on the angular quantum number \(l\) and the coupling coefficient \(\zeta\). To analyze stability, we introduce the dimensionless coordinate \(x=r/r_{0} > 1\). The effective potential (\ref{e13}) can be rewritten as:
\begin{equation}
\tilde{V}(x) = \left(1 - \frac{1}{x^4}\right) \left[ \frac{l(l+1)}{x^2} + \frac{4 + 6\zeta}{x^6} \right].
\end{equation}
Since the prefactor is positive for \(x > 1\), the sign of the potential is determined by the term in the brackets. For the monopole perturbation (\(l=0\)), the condition for the potential to develop a negative region near the horizon (\(x \to 1\)) is:
\begin{equation}
4 + 6\zeta < 0 \implies \zeta < -\frac{2}{3}.
\end{equation}

From a physical perspective, a negative non-minimal coupling (\(\zeta < 0\)) indicates that the effective curvature acts as an attractive potential for the scalar field. In the context of curved spacetime quantum field theory and cosmology, such negative coupling often leads to a negative effective mass squared (\(m_{\text{eff}}^2 \propto \zeta R\)), which is the hallmark of a tachyonic instability. This mechanism is widely used to trigger spontaneous symmetry breaking or cosmic inflation. In our analogue gravity system, the emergence of the negative well for \(\zeta < -2/3\) suggests that the effective acoustic geometry may no longer support stable wave propagation, potentially resulting in an exponential energy accumulation. It should be emphasized that the presence of a negative potential region is a necessary but not sufficient condition for instability: whether a genuine exponentially growing mode actually develops depends on the depth and width of the well, on the boundary conditions, and on whether a normalizable unstable mode is supported. Therefore, the threshold \(\zeta_{\text{crit}} = -2/3\) should be interpreted as a change in the qualitative behavior of the effective potential, which may signal a dynamical transition from a stable acoustic spacetime to a potentially tachyonically unstable one.

Therefore, \(\zeta_{\text{crit}} = -2/3\) serves as an analytical bound. For \(\zeta \ge -2/3\), the effective potential is strictly positive, and no tachyonic instability of this type is expected. For \(\zeta < -2/3\), a negative potential well emerges near the horizon, which may give rise to a tachyonic instability.

\begin{figure}[h]
\centering
\includegraphics[width=0.6\textwidth]{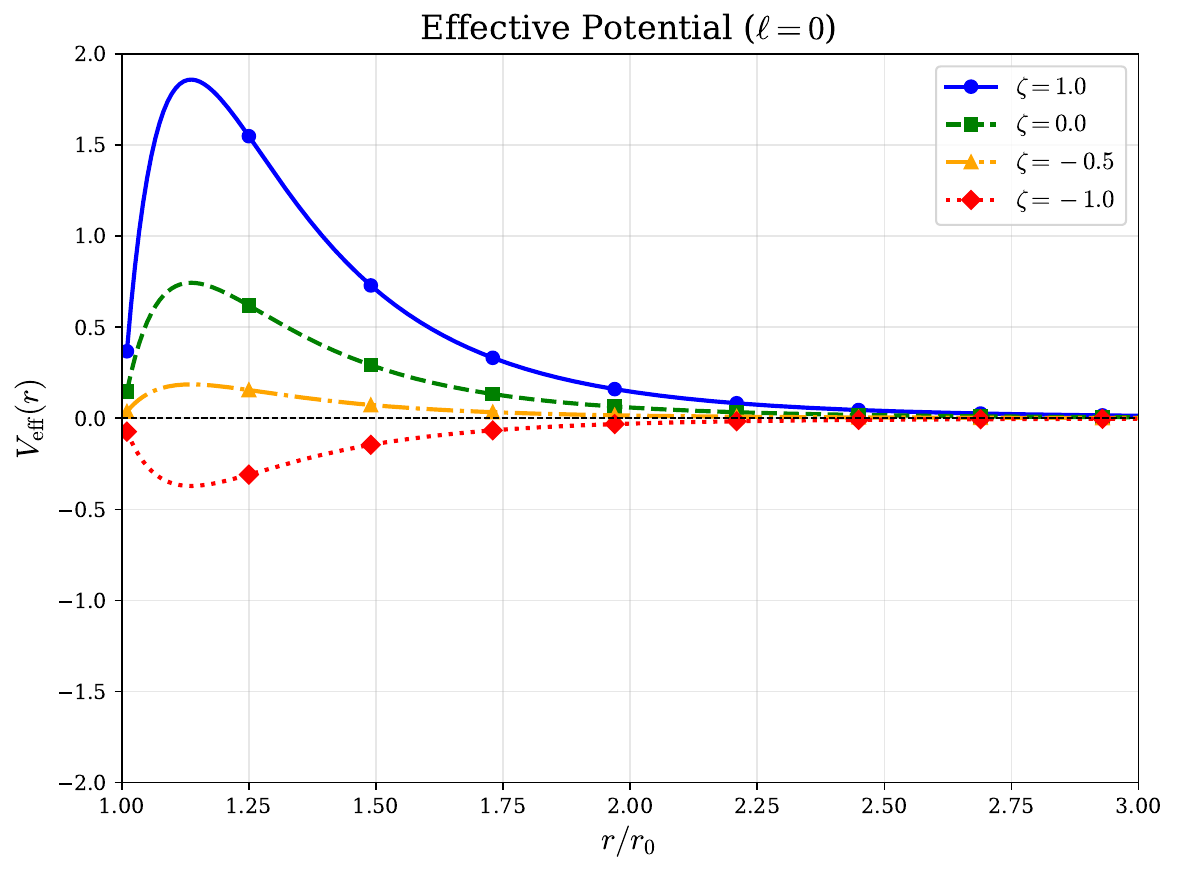} 
\caption{The effective potential \(V(r)\) for \(l=0\) with different values of the coupling parameter \(\zeta\). A negative potential well emerges for \(\zeta = -1.0 < \zeta_{\text{crit}}\), which may indicate a tachyonic instability.}
\label{f1}
\end{figure}

We have plotted the effective potential \(V(r)\) as a function of \(r\) for \(l=0\) in Fig. \ref{f1} for several values of \(\zeta\). As clearly shown in the figure, while the potential remains positive for \(\zeta = 1.0, 0.0\) and \(-0.5\), a deep negative well develops for \(\zeta = -1.0\) (which is less than the critical value \(\zeta_{\text{crit}} = -2/3\)). Throughout the rest of this paper, we restrict our QNM analysis to the stable regime \(\zeta \ge -2/3\).

It should be noted that the critical threshold \(\zeta_{\text{crit}} = -2/3\) is strictly derived for the monopole perturbation (\(l=0\)). For higher multipole numbers (\(l \ge 1\)), the centrifugal term \(l(l+1)/x^2\) in Eq. (\ref{e13}) provides a positive contribution near the horizon, which competes with the negative contribution from the coupling term. Consequently, for \(l \ge 1\), the development of a negative potential well requires a more negative coupling parameter, i.e., \(\zeta < -2/3\). Because this negative region only appears locally for \(l \ge 1\) and the potential barrier geometry becomes more complex, the WKB approximation is not suitable for locating the exact threshold in this regime. Therefore, in this work, the analytical bound \(\zeta_{\text{crit}} = -2/3\) serves as the most stringent (conservative) condition for the onset of a negative potential region, and our subsequent QNM analysis and time-domain simulations focus on the \(l=0\) case to demonstrate this behavior.

\begin{figure}[htbp]
\centering
\includegraphics[width=0.48\textwidth]{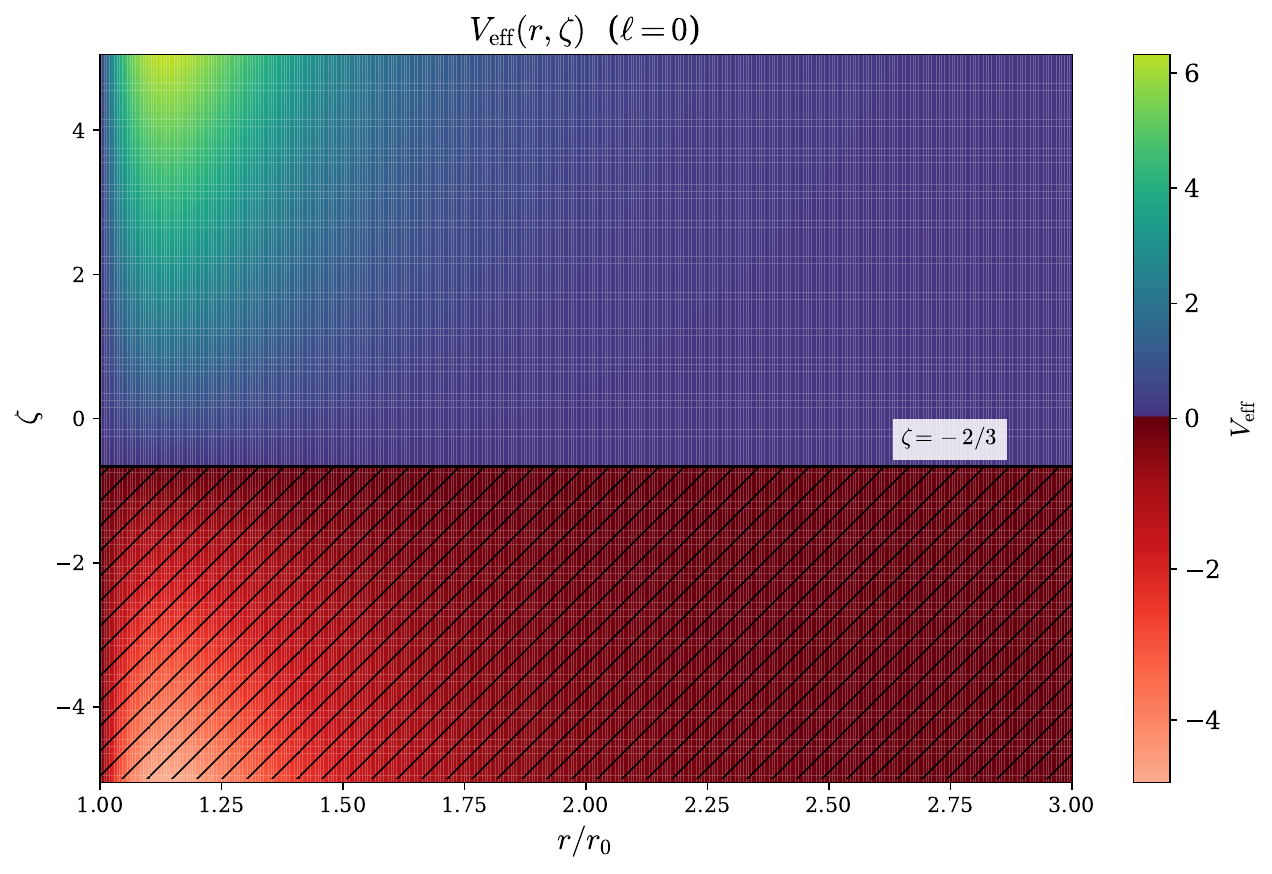} 
\hfill
\includegraphics[width=0.48\textwidth]{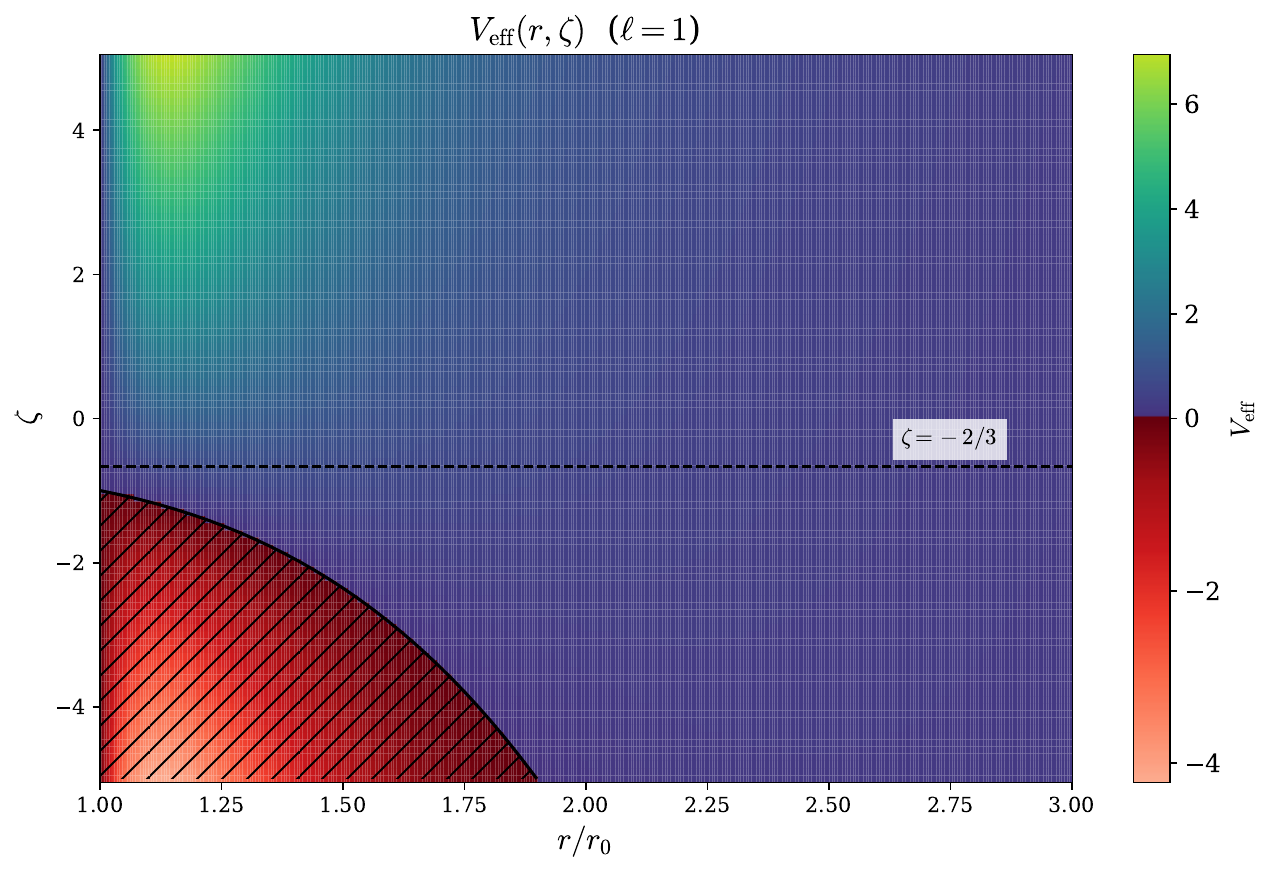}
\caption{The effective potential \(V_{\mathrm{eff}}(r,\zeta)\) in the \((r,\zeta)\) plane. 
(a) For the monopole perturbation (\(\ell=0\)), the negative region (\(V_{\mathrm{eff}}<0\), red) lies strictly below the critical threshold \(\zeta=-2/3\) (white dashed line). 
(b) For \(\ell=1\), the red region is pushed to more negative values of \(\zeta\) due to the centrifugal barrier. The black hatched region (\(V_{\mathrm{eff}}<0\)) ensures clear visibility in grayscale printing.}
\label{fig:heatmaps}
\end{figure}

To visualize the parameter space of stability, we present the effective potential \(V_{\mathrm{eff}}(r,\zeta)\) in the \((r,\zeta)\) plane for \(\ell=0\) and \(\ell=1\) in Fig. \ref{fig:heatmaps}. As shown in Fig. \ref{fig:heatmaps}(a), for the monopole case (\(\ell=0\)), the parameter space is sharply divided by the \(\zeta=-2/3\) boundary: the region with \(\zeta < -2/3\) is entirely dominated by negative values (shown in red and hatched). For \(\ell=1\) [Fig. \ref{fig:heatmaps}(b)], the centrifugal term \(l(l+1)/x^2\) counteracts the negative contribution from the coupling. Consequently, the region with negative potential is compressed into a smaller corner near the horizon, requiring a more negative coupling (approximately \(\zeta < -1\) at \(r \to r_0\)) to trigger a negative potential.

\begin{figure}[htbp]
\centering
\includegraphics[width=0.7\textwidth]{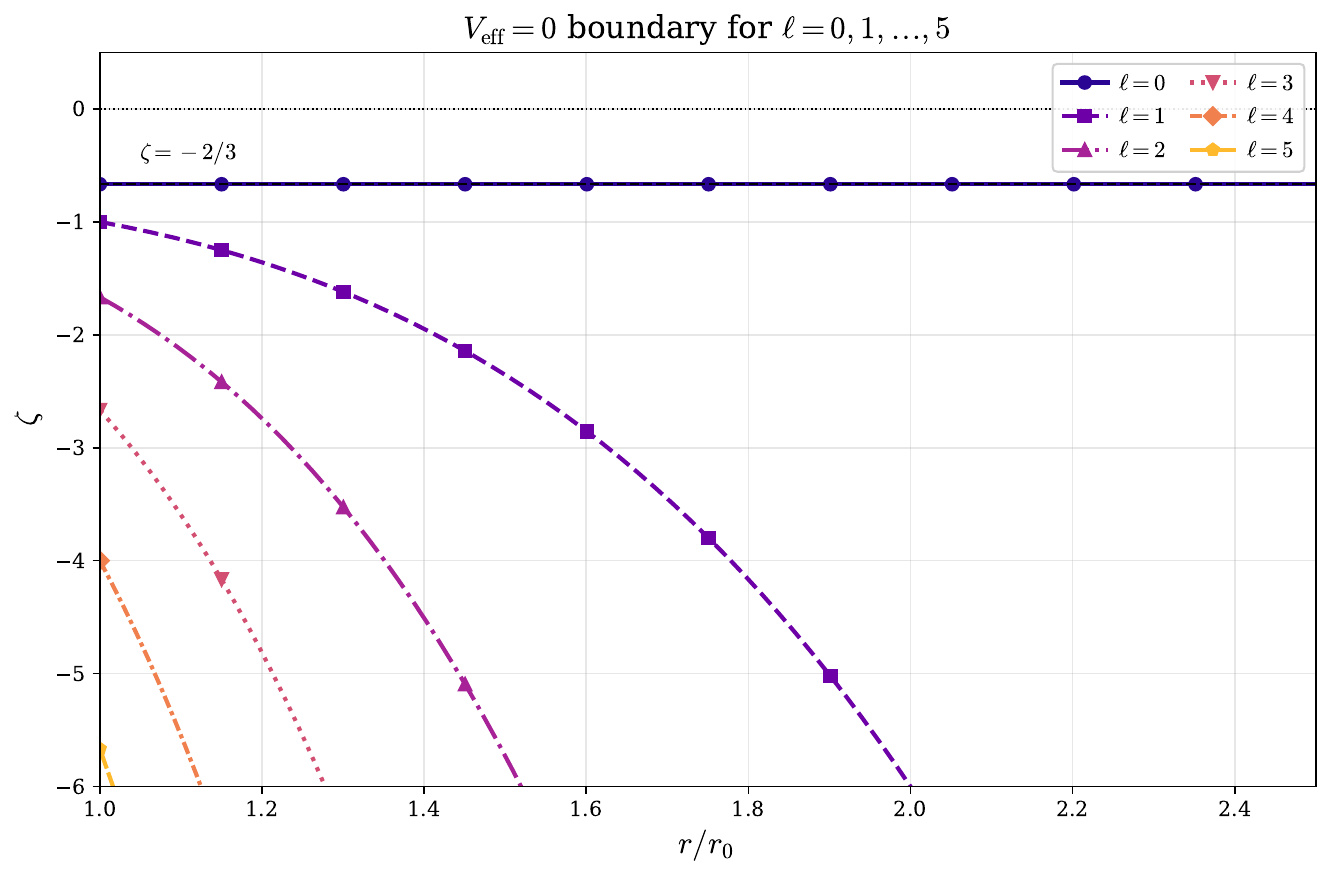}
\caption{The analytical \(V_{\mathrm{eff}}=0\) boundary curves in the \((r,\zeta)\) plane for multipole numbers \(\ell=0,1,\dots,5\). As \(\ell\) increases, the boundary shifts downward toward more negative \(\zeta\), demonstrating that the monopole case (\(\ell=0\)) provides the most stringent bound at \(\zeta_{\rm crit}=-2/3\) for the absence of a negative potential region. Curves are distinguished by different line styles and markers to ensure clarity in grayscale.}
\label{fig:boundaries}
\end{figure}

To further generalize this behavior, we analytically solve \(V_{\mathrm{eff}}=0\) to obtain the boundary curve \(\zeta(r) = -[\ell(\ell+1)r^4+4]/6\), which is plotted in Fig. \ref{fig:boundaries} for \(\ell=0,1,\dots,5\). For the monopole case (\(\ell=0\)), the boundary is the horizontal line \(\zeta=-2/3\). As \(\ell\) increases, the boundary curve shifts downward toward more negative values of \(\zeta\). This confirms that the threshold \(\zeta_{\rm crit}=-2/3\) derived for the monopole case provides the most stringent (conservative) condition for the effective potential to remain non-negative: if \(\zeta \ge -2/3\), the potential is non-negative for all \(\ell\) and all \(r > r_0\). Conversely, if \(\zeta < -2/3\), only the monopole perturbation can necessarily develop a negative potential region, and higher multipoles require even more negative coupling to trigger the instability.

\subsection{QNMs of scalar perturbation}
In this work, we compute the quasinormal mode (QNM) frequencies of the non-minimally coupled scalar field in the canonical acoustic black hole background using the sixth-order WKB approximation. The results for multipole numbers \(l=4\) and \(l=5\) are listed in Tables \ref{tb1} and \ref{tb2}, respectively, across various coupling parameters \(\zeta\) and overtone numbers \(n\).

\begin{table}[th]
\centering
\begin{tabular}{c|c|c|c|c|c}
\hline
\(\zeta\) & \(\omega (n=0)\) & \(\omega (n=1)\) & \(\omega (n=2)\) & \(\omega (n=3)\) & \(\omega (n=4)\) \\ \hline
\(-0.5\) & \(2.68263-0.60777\mathrm{i}\) & \(2.36893-1.86796\mathrm{i}\) & \(1.69801-3.33030\mathrm{i}\) & \(0.70627-5.27028\mathrm{i}\) & \(0.47842+7.94752\mathrm{i}^{\ast}\) \\ \hline
\(0.0\)  & \(2.75319-0.61870\mathrm{i}\) & \(2.43969-1.90336\mathrm{i}\) & \(1.77142-3.39678\mathrm{i}\) & \(0.78650-5.36918\mathrm{i}\) & \(0.39336+8.07216\mathrm{i}^{\ast}\) \\ \hline
\(0.5\)  & \(2.82380-0.63224\mathrm{i}\) & \(2.50419-1.95338\mathrm{i}\) & \(1.83062-3.49779\mathrm{i}\) & \(0.83174-5.51074\mathrm{i}\) & \(0.40320+8.21713\mathrm{i}^{\ast}\) \\ \hline
\(1.0\)  & \(2.89406-0.64772\mathrm{i}\) & \(2.56726-2.00697\mathrm{i}\) & \(1.87817-3.58894\mathrm{i}\) & \(0.81435-5.64004\mathrm{i}\) & \(0.56013+8.42056\mathrm{i}^{\ast}\) \\ \hline
\end{tabular}
\caption{The QNM frequencies (\(\omega = \omega_R + i\omega_I\)) of the non-minimally coupled scalar field for \(l=4\), computed via the sixth-order WKB approximation across different values of \(\zeta\) and \(n\). Entries marked with \(\ast\) (where \(n\ge l\)) lie outside the WKB validity regime due to the positive imaginary part; they are shown only for completeness and are excluded from the physical analysis.}
\label{tb1}
\end{table}

\begin{table}[th]
\centering
\begin{tabular}{c|c|c|c|c|c|c}
\hline
\(\zeta\) & \(\omega (n=0)\) & \(\omega (n=1)\) & \(\omega (n=2)\) & \(\omega (n=3)\) & \(\omega (n=4)\) & \(\omega (n=5)\) \\ \hline
\(-0.5\) & \(3.32295-0.61267\mathrm{i}\) & \(3.06784-1.86750\mathrm{i}\) & \(2.52950-3.24031\mathrm{i}\) & \(1.69975-4.89322\mathrm{i}\) & \(0.64384-7.01631\mathrm{i}\) & \(0.55398+9.77207\mathrm{i}^{\ast}\) \\ \hline
\(0.0\)  & \(3.38007-0.61989\mathrm{i}\) & \(3.12505-1.88954\mathrm{i}\) & \(2.58753-3.27811\mathrm{i}\) & \(1.76010-4.94683\mathrm{i}\) & \(0.70727-7.08455\mathrm{i}\) & \(0.48737+9.85349\mathrm{i}^{\ast}\) \\ \hline
\(0.5\)  & \(3.43804-0.62794\mathrm{i}\) & \(3.18141-1.91497\mathrm{i}\) & \(2.64115-3.32427\mathrm{i}\) & \(1.81095-5.01476\mathrm{i}\) & \(0.75231-7.16813\mathrm{i}\) & \(0.45688+9.93995\mathrm{i}^{\ast}\) \\ \hline
\(1.0\)  & \(3.49652-0.63683\mathrm{i}\) & \(3.23666-1.94404\mathrm{i}\) & \(2.69031-3.37802\mathrm{i}\) & \(1.84867-5.09146\mathrm{i}\) & \(0.76255-7.25455\mathrm{i}\) & \(0.50264+10.01192\mathrm{i}^{\ast}\) \\ \hline
\end{tabular}
\caption{Same as Table \ref{tb1}, but for \(l=5\) and overtone numbers up to \(n=5\).}
\label{tb2}
\end{table}
To validate our numerical implementation, we perform a detailed comparison with previous studies.

First, we compare our results with the sixth-order WKB calculations of Matyjasek et al.\cite{Maty}. For the \((3+1)\)-dimensional background and angular quantum numbers \(l \ge 1\), our fundamental mode results match their reported values exactly, with all displayed digits being identical. However, for the extremal case of \(l=0\), they reported \(0.05820 - 0.87431\mathrm{i}\), whereas our calculation yields \(0.05820 + 0.8743\mathrm{i}\).The discrepancy in the sign of the imaginary part is a well-known artifact of the WKB approximation. For \(l=0\), the effective potential barrier becomes extremely flat and broad, violating the assumptions required for the WKB method to be accurate. In such regimes, the WKB series breaks down and can yield positive imaginary parts, reflecting an unphysical instability rather than a true damped mode. This behavior is perfectly consistent with our own numerical results (see Tables \ref{tb1} and \ref{tb2}): the WKB approximation breaks down and produces positive imaginary parts precisely when the overtone number satisfies \(n \ge l\) (for instance, the positive imaginary parts appear at \(l=4, n=4\) and \(l=0, n=0\)). 

This naturally explains why WKB-based QNM studies are typically restricted to the low-overtone regime (\(n < l\)), where the effective potential barrier is well-defined and the WKB series remains accurate. Our restriction to this reliable regime therefore ensures the physical validity of the baseline analysis.

Second, we compare our results with the third-order WKB calculations of Chen et al. \cite{Chen}. Their work focused on \(l=5\) with coupling parameters ranging from \(\xi=0\) to \(5\). For the fundamental mode and lower overtones, our results are in close agreement with theirs, showing only minor deviations. As the overtone number \(n\) increases, however, the discrepancies become more pronounced. This behavior is consistent with the asymptotic nature of the WKB series, where higher-order corrections are essential for accurately capturing highly excited modes, while lower-order approximations lose accuracy at high overtones. 

The exact agreement with Matyjasek et al. for \(l \ge 1\), the expected sign flip for the flat potential at \(l=0\), and the expected deviation from the third-order results at high overtones together justify our use of the sixth-order WKB approximation for the baseline analysis in this work.

 A detailed analysis of the data in Tables \ref{tb1} and \ref{tb2} reveals several clear physical trends.

\subsubsection*{1. Dependence on the coupling parameter \(\zeta\)}
For all reliable modes, both the real frequency \(\omega_R\) and the damping rate \(|\omega_I|\) exhibit a monotonic increase with the coupling parameter \(\zeta\). Taking the fundamental modes in Table \ref{tb1} as an example, \(\omega_R\) increases from \(2.68263\) to \(2.89406\), and \(|\omega_I|\) increases from \(0.60777\) to \(0.64772\) as \(\zeta\) varies from \(-0.5\) to \(1.0\). Physically, a larger \(\zeta\) enhances the repulsive effective potential barrier near the horizon, compressing the wave function and leading to a higher oscillation frequency and faster decay. The only exception is the highly excited mode (\(l=4, n=3\)) in Table \ref{tb1}, where \(\omega_R\) drops slightly from \(0.83174\) to \(0.81435\). This minor deviation arises because this mode lies close to the validity boundary of the WKB approximation, making it sensitive to parameter variations.

\subsubsection*{2. Dependence on the overtone number \(n\) and WKB breakdown}
For low-lying overtones, \(\omega_R\) increases with the overtone number \(n\). However, the tables demonstrate a clear breakdown of the WKB approximation when \(n \ge l\): the imaginary part of the frequency abruptly becomes positive (e.g., \(\omega_I = +7.94752\) at \(l=4, n=4, \zeta=0\)), implying an unphysical exponentially growing mode. This anomaly occurs because, as \(n\) approaches or exceeds \(l\), the effective potential barrier becomes extremely broad and flat, violating the single-peaked assumption of the WKB method. Consequently, data for \(n \ge l\) fall outside the regime of physical validity and are excluded from our analysis. This also naturally explains why WKB-based QNM studies typically restrict their focus to the low-overtone regime (\(n < l\)).

It is important to note that Table \ref{tb1} only lists the QNM frequencies in the stable regime (\(\zeta \ge -2/3\)). For \(\zeta < \zeta_{\text{crit}} = -2/3\), the emergence of a negative potential well introduces a second turning point. This classically allowed region fundamentally violates the single-barrier assumption of the WKB formula, yielding spurious results. Therefore, to rigorously examine whether this negative well does indeed induce a tachyonic instability, we rely exclusively on the time-domain evolution presented in the following section.

\begin{figure}[htbp]
\centering
\includegraphics[width=0.48\textwidth]{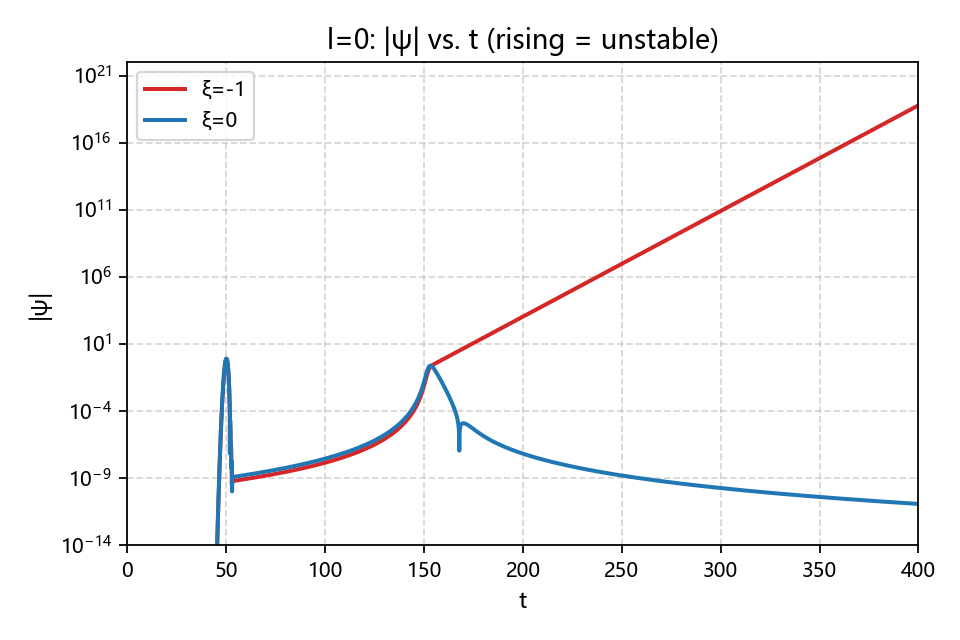}
\includegraphics[width=0.48\textwidth]{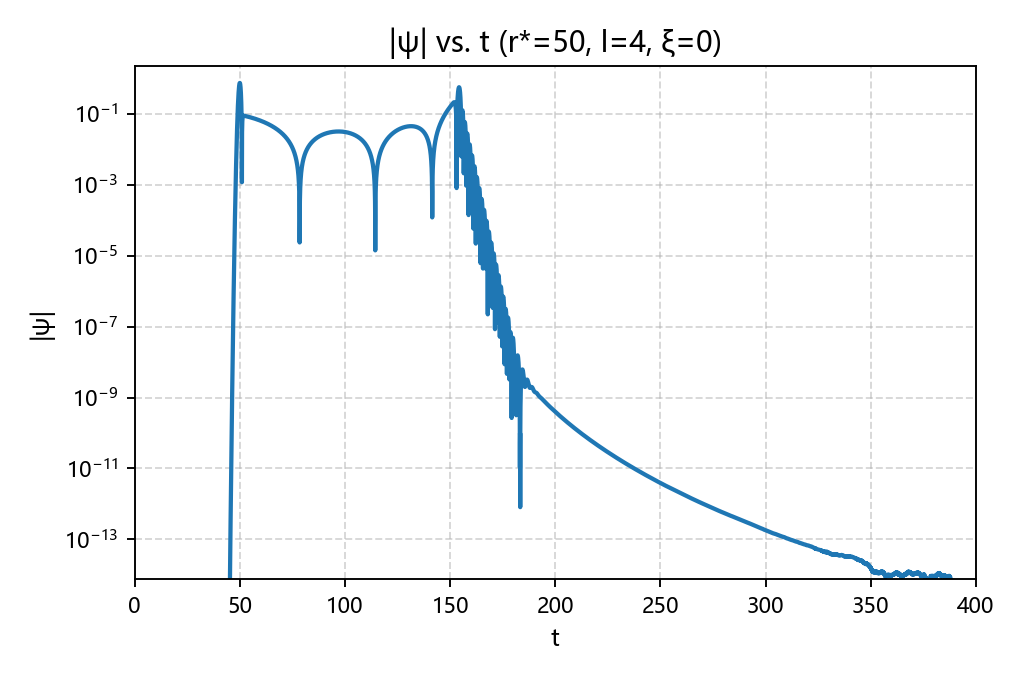}
\caption{Time-domain evolution of the perturbation amplitude \(|\psi|\) on a logarithmic scale over \(t \in [0, 400]\). (a) \(l=0\) for \(\zeta=-1.0\) (unstable, exponential growth) and \(\zeta=0.0\) (stable, decay). (b) \(l=4\) and \(\zeta=0.0\), observed at \(r_*=50\), showing the damped ringdown oscillations after the initial pulse.}
\label{f_time}
\end{figure}

To further examine whether the negative potential well does indeed lead to a tachyonic instability for \(\zeta < -2/3\), we performed a time-domain evolution of the scalar field perturbation. In the tortoise coordinate \(r_*\), wave equation (\ref{e6}) can be recast into the following first-order form:
\begin{equation}
\partial_t \psi = \Pi, \quad \partial_t \Pi = \partial_{r_*}^2 \psi - V(r_*)\psi,
\end{equation}
where \(\psi\) is the perturbation amplitude and \(\Pi\) is its conjugate momentum. We solve this system using the standard method of lines with a fourth-order Runge-Kutta integrator. The spatial derivatives are evaluated using second-order central finite differences on a uniform grid in \(r_*\). An initial Gaussian wave packet is placed far from the horizon, and approximate absorbing (outgoing-wave) boundary conditions are imposed at both the horizon and spatial infinity to suppress non-physical reflections. The small residual reflections visible at late times do not affect the fitted exponential growth rate in the unstable case. The results are shown in Fig. \ref{f_time}.

As clearly depicted in Fig. \ref{f_time}, for the minimally coupled case (\(\zeta = 0.0\)), the perturbation decays rapidly after the initial ringdown phase. However, for \(\zeta = -1.0\) (which is below the critical threshold \(\zeta_{\text{crit}} = -2/3\)), the amplitude grows exponentially with time, manifesting as a straight line with a positive slope in the logarithmic plot. Although the late-time numerical oscillations in the stable case are slightly affected by boundary reflections, this time-domain simulation provides a robust qualitative confirmation of the analytical prediction of the tachyonic instability.

\section{Electromagnetic field perturbation of acoustic black hole}

Having examined the non-minimally coupled scalar sector and identified its critical threshold (\(\zeta_{\text{crit}} = -2/3\)), a natural question arises: is this behavior a universal feature of the acoustic black hole geometry, or is it exclusively induced by the non-minimal coupling? To answer this, we now turn our attention to the electromagnetic and Dirac fields. Since these fields are governed by gauge invariance and minimal coupling, they serve as ideal baselines to isolate the curvature coupling effects. In this section, we investigate the stability of the canonical acoustic black hole under the perturbation of the electromagnetic field. Here we consider the electromagnetic field exterior to the black hole, its evolution is governed by Maxwell's equation
\begin{equation}
F_{;\upsilon }^{\mu \upsilon }=0,  \label{b1}
\end{equation}
where $F^{\mu \nu }$ is the electromagnetic field strength defined via the electromagnetic potential $A_{\mu }$
\begin{equation}
F_{\mu \upsilon }=A_{\upsilon ,\mu }-A_{\mu ,\nu }.  \label{b2}
\end{equation}

Due to the spherical symmetry of the convergent acoustic black hole, we take the form of the electromagnetic potential as
\begin{equation}
A_{t}=A_{r}=A_{\theta }=0,A_{\phi }=\psi \left( r,t\right) \sin \theta \frac{dP_{l}\left( \cos \theta \right) }{d\theta },  \label{b3}
\end{equation}
where \(P_{l}\left( \cos \theta \right) \) is the Legendre function.

From Eqs. (\ref{b2}) and (\ref{b3}), we obtain the nonvanishing components of the electromagnetic tensor. Substituting these into Eq. (\ref{b1}), we get 
\begin{eqnarray}
&&\frac{1}{r^{2}\sin \theta }\frac{dP_{l}\left( \cos \theta \right) }{%
d\theta }\frac{\partial ^{2}\psi \left( r,t\right) }{\partial t^{2}}-\frac{1%
}{r^{2}\sin \theta }\frac{dP_{l}\left( \cos \theta \right) }{d\theta }\left[
f\left( r\right) \frac{\partial ^{2}\psi \left( r,t\right) }{\partial r^{2}}+%
\frac{df\left( r\right) }{dr}\frac{\partial \psi \left( r,t\right) }{%
\partial r}\right]  \label{b10} \\
&&+\frac{1}{r^{4}\sin \theta }\psi \left( r,t\right) \left[ \csc ^{2}\theta 
\frac{dP_{l}\left( \cos \theta \right) }{d\theta }-\cot\theta \frac{%
d^{2}P_{l}\left( \cos \theta \right) }{d\theta ^{2}}-\frac{d^{3}P_{l}\left(
\cos \theta \right) }{d\theta ^{3}}\right] =0.  \notag
\end{eqnarray}

Since the Legendre function satisfies

\begin{equation}
\csc ^{2}\theta \frac{dP_{l}\left( \cos \theta \right) }{d\theta }-\cot\theta 
\frac{d^{2}P_{l}\left( \cos \theta \right) }{d\theta ^{2}}-\frac{%
d^{3}P_{l}\left( \cos \theta \right) }{d\theta ^3}=l\left( l+1\right) \frac{%
dP_{l}\left( \cos \theta \right) }{d\theta },  \label{b11}
\end{equation}

so Eq. (\ref{b10}) can be abbreviated as

\begin{equation}
\frac{\partial ^{2}\psi \left( r,t\right) }{\partial t^{2}}-\left[ f\left(
r\right) \frac{\partial ^{2}\psi \left( r,t\right) }{\partial r^{2}}+\frac{%
df\left( r\right) }{dr}\frac{\partial \psi \left( r,t\right) }{\partial r}%
\right] +\frac{l\left( l+1\right) }{r^{2}}\psi \left( r,t\right) =0.
\label{b12}
\end{equation}

\bigskip Introduce a tortoise transformation as Eq. (\ref{e8}), and
decompose $\psi \left( r,t\right) $\ as

\begin{equation}
\psi \left( r,t\right) =\frac{\varphi \left( r\right) }{r}e^{-i\omega t},
\label{b13}
\end{equation}

then, substitute into Eq. (\ref{b12}), we get the Schr\"{o}dinger-like
equation

\begin{equation}
\frac{d^{2}\varphi (r)}{dr_{\ast }^{2}}+\left( \omega ^{2}-V\left( r\right)
\right) \varphi (r)=0,  \label{b14}
\end{equation}

the effective potential $V\left( r\right) $\ reads

\begin{equation}
V\left( r\right) =\frac{l\left( l+1\right) }{r^{2}}f\left( r\right) .
\label{b15}
\end{equation}

\bigskip We can find that the effective potential above is relevant to the
angular quantum number $l$. The form of the effective potential is similar
to that of the Schwarzschild black hole under the perturbation of
electromagnetic field\cite{KK}.

\subsection{Effective potential and QNMs of electromagnetic field perturbation}

\begin{figure}[h]
\centering
\includegraphics[width=0.7\linewidth]{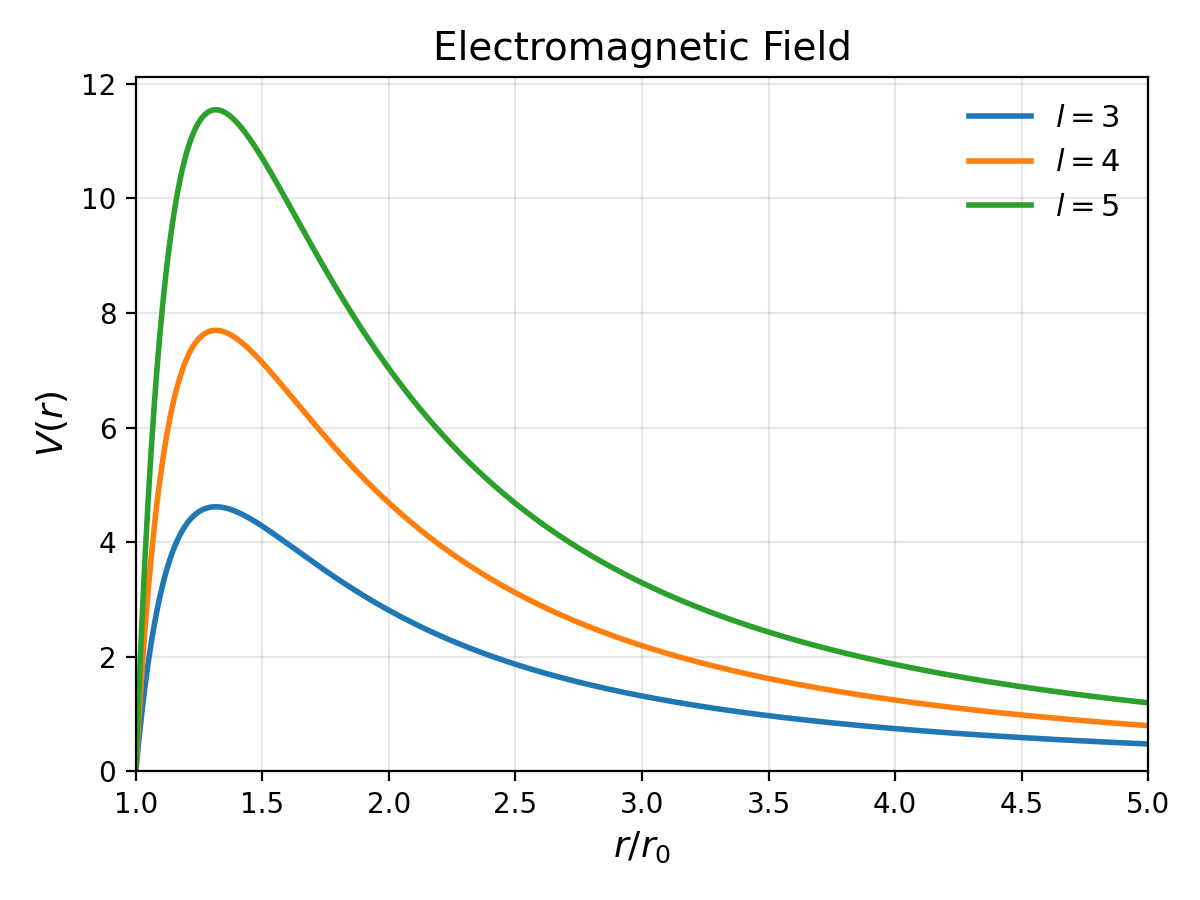}
\caption{The effective potential \(V(r)\) for the electromagnetic field perturbation with \(l=3,4,5\). The radial coordinate is normalized as \(r/r_0\). The potential is purely geometric and independent of the coupling parameter \(\zeta\); it serves as the minimally coupled baseline.}
\label{fig:em_potential}
\end{figure}

For better visualization, we plot \(V(r)\) versus \(r\) for \(l=3,4,5\) in Figure \ref{fig:em_potential}. From the figure, we can find that both the value of the effective potential and the position of its peak grow with \(l\). Importantly, as seen from Eq. (\ref{b15}), the effective potential for the electromagnetic field is purely geometric and does not contain coupling with the Ricci scalar \(R\). This is physically expected due to the gauge invariance of the electromagnetic field. Consequently, the QNMs of the electromagnetic field are independent of the parameter \(\zeta\) and serve as a \textbf{minimally coupled baseline}.

Just like in the case of a scalar field perturbation, we evaluate the QNMs of the electromagnetic field using the sixth-order WKB approximation. The results are listed in Table \ref{tb3}.

\begin{table}[h]
	\centering
	\begin{tabular}{c|c|c|c|c|c}
		\hline
		$l$ & $n$ & $\omega$ & $l$ & $n$ & $\omega$ \\ \hline
		4 & 0 & $2.65952-0.60448\mathrm{i}$ & 5 & 0 & $3.30416-0.61044\mathrm{i}$
		\\ \hline
		& 1 & $2.34484-1.85770\mathrm{i}$ &  & 1 & $3.04869-1.86063\mathrm{i}$ \\ 
		\hline
		& 2 & $1.67119-3.31250\mathrm{i}$ &  & 2 & $2.50929-3.22845\mathrm{i}$ \\ 
		\hline
		& 3 & $0.67512-5.24482\mathrm{i}$ &  & 3 & $1.67749-4.87625\mathrm{i}$ \\ 
		\hline
		& 4 &  &  & 4 & $0.61886-6.99411\mathrm{i}$ \\ 
		\hline
	\end{tabular}%
	\caption{The quasi normal mode spectrum of acoustic black hole under the
		electromagnetic field perturbations (angular quantum number $l=4,5$). Blank entries correspond to modes for which the sixth-order WKB approximation breaks down (positive imaginary part), and are therefore omitted.}
	\label{tb3}
\end{table}

From Table \ref{tb3}, we can conclude that the real part of \(\omega_n\) decreases with \(n\), while the absolute value of the imaginary part increases. For the fundamental mode and the lower overtones, the imaginary part is strictly negative, which means that the vibration decays, and the canonical acoustic black hole is stable against the perturbation of the electromagnetic field. 

\section{Dirac field perturbation to the canonical acoustic black hole}

Similarly to the electromagnetic case, the Dirac field is also minimally coupled. We now analyze its perturbation to further confirm the stability of the minimally coupled sector.

\subsection{The effective potential of the Dirac field perturbation}
The dynamics of a massless spin-1/2 field in curved spacetime is described
by the Dirac equation\cite{Brill}

\begin{equation}
\left[ \gamma ^{a}e_{a}^{\mu }\left( \partial _{\mu }+\Gamma _{\mu }\right) %
\right] \Psi =0,  \label{c1}
\end{equation}%
where $e_{a}^{\mu }$ is the inverse of the tetrad $e_{\mu }^{a}$ with $%
g_{\mu \nu }=\eta _{ab}e_{\mu }^{a}e_{\nu }^{b}$, $\eta _{ab}$ is the
Minkowski metric. $\gamma ^{a}$ are the Dirac matrices, defined as

\begin{equation}
\gamma ^{0}=\left( 
\begin{array}{cc}
-i & 0 \\ 
0 & i%
\end{array}%
\right) ,\gamma ^{j}=\left( 
\begin{array}{cc}
0 & -i\sigma ^{j} \\ 
i\sigma ^{j} & 0%
\end{array}%
\right) \left( j=1,2,3\right) ,  \label{c2}
\end{equation}%
\ \ $\sigma ^{j}$ is the Pauli matrices, defined as $\sigma ^{1}=%
\begin{pmatrix}
0 & 1 \\ 
1 & 0%
\end{pmatrix}%
$, $\sigma ^{2}=%
\begin{pmatrix}
0 & -i \\ 
i & 0%
\end{pmatrix}%
$, $\sigma ^{3}=%
\begin{pmatrix}
1 & 0 \\ 
0 & -1%
\end{pmatrix}%
$. And the spin connections $\Gamma _{\mu }$ are given by

\begin{equation}
\Gamma _{\mu }=\frac{1}{8}\left[ \gamma ^{a},\gamma ^{b}\right] e_{a}^{\nu
}e_{b\nu ;\mu },\quad e_{b\nu ;\mu }=\partial _{\mu }e_{b\nu }-\Gamma _{\mu
\nu }^{\alpha }e_{b\alpha }.  \label{c3}
\end{equation}

\bigskip To separate the Dirac equation, we choose the tetrad

\begin{equation}
e_{\mu }^{a}=diag\left( \sqrt{f},\frac{1}{\sqrt{f}},r,r\sin \theta \right) ,
\label{c4}
\end{equation}%
and substitute this tetrad into Eq. (\ref{c1}). Then the Dirac equation
becomes

\begin{equation}
\frac{\gamma ^{0}}{\sqrt{f}}\frac{\partial \psi }{\partial t}+\sqrt{f}\gamma
^{1}\left( \frac{\partial }{\partial r}+\frac{1}{r}+\frac{1}{4f}\frac{df}{dr}
\right) \psi +\frac{\gamma ^{2}}{r}\left( \frac{\partial }{\partial \theta }%
+ \frac{1}{2}\cot \theta \right) \psi +\frac{\gamma ^{3}}{r\sin \theta }%
\frac{ \partial \psi }{\partial \varphi }=0.  \label{c5}
\end{equation}

Defining the rescaled perturbation $\psi =f^{-\frac{1}{4}}\phi $, the
equation above becomes

\begin{equation}
\frac{\gamma ^{0}}{\sqrt{f}}\frac{\partial \phi }{\partial t}+\sqrt{f}\gamma
^{1}\left( \frac{\partial }{\partial r}+\frac{1}{r}\right) \phi +\frac{%
\gamma ^{2}}{r}\left( \frac{\partial }{\partial \theta }+\frac{1}{2}\cot
\theta \right) \phi +\frac{\gamma ^{3}}{r\sin \theta }\frac{\partial \phi }{%
\partial \varphi }=0.  \label{c6}
\end{equation}

\bigskip \bigskip Introducing the tortoise coordinate 
\begin{equation}
r_{\ast }=\int \frac{\mathrm{d}r}{f\left( r\right) },  \label{c7}
\end{equation}%
and the ansatz for the Dirac spinor

\begin{equation}
\phi \left( t,r,\theta ,\varphi \right) =%
\begin{pmatrix}
i\frac{G^{\left( \pm \right) }\left( r\right) \chi _{jm}^{\pm }\left( \theta
,\varphi \right) }{r} \\ 
\frac{F^{\left( \pm \right) }\left( r\right) \chi _{jm}^{\mp }\left( \theta
,\varphi \right) }{r}%
\end{pmatrix}%
e^{-i\omega t},  \label{c8}
\end{equation}%
where $\chi _{jm}^{\pm }$ are spinning spherical harmonics given by 
\begin{equation}
\chi _{jm}^{+}=%
\begin{pmatrix}
\sqrt{\frac{j+m}{2j}}Y_{\ell }^{m-\frac{1}{2}}\left( \theta ,\varphi \right)
\\ 
\sqrt{\frac{j-m}{2j}}Y_{\ell }^{m+\frac{1}{2}}\left( \theta ,\varphi \right)%
\end{pmatrix}%
,j=\ell +\frac{1}{2},  \label{c9}
\end{equation}%
and 
\begin{equation}
\chi _{jm}^{-}=%
\begin{pmatrix}
\sqrt{\frac{j-m+1}{2\left( j+1\right) }}Y_{l}^{m-\frac{1}{2}}\left( \theta
,\varphi \right) \\ 
\sqrt{\frac{j+m+1}{2\left( j+1\right) }}Y_{l}^{m+\frac{1}{2}}\left( \theta
,\varphi \right)%
\end{pmatrix}%
,j=\ell -\frac{1}{2},  \label{c10}
\end{equation}%
where$\ Y_{l}^{m\pm \frac{1}{2}}\left( \theta ,\varphi \right) $ are the
usual spin-weighted spherical harmonics. \bigskip Thus, Eq. (\ref{c6}) can
be written in a simplified manner as: 
\begin{equation}
\begin{pmatrix}
0 & -\omega \\ 
\omega & 0%
\end{pmatrix}%
\begin{pmatrix}
F^{\pm } \\ 
G^{\pm }%
\end{pmatrix}%
-\frac{\partial }{\partial r_{\ast }}%
\begin{pmatrix}
F^{\pm } \\ 
G^{\pm }%
\end{pmatrix}%
+\sqrt{f}%
\begin{pmatrix}
\frac{k_{\pm }}{r} & 0 \\ 
0 & -\frac{k_{\pm }}{r}%
\end{pmatrix}%
\begin{pmatrix}
F^{\pm } \\ 
G^{\pm }%
\end{pmatrix}%
=0.  \label{c11}
\end{equation}%
The cases (+) and (-) in the functions can be put together after some
matching, and the equation can be decoupled as

\begin{equation}
\frac{\mathrm{d^{2}}F}{\mathrm{d}r_{\ast }^{2}}+\left( \omega
^{2}-V_{1}\right) F=0,  \label{c12}
\end{equation}

\begin{equation}
\frac{\mathrm{d^{2}}G}{\mathrm{d}r_{\ast }^{2}}+\left( \omega
^{2}-V_{2}\right) G=0,  \label{c13}
\end{equation}%
where effective potential $V_{1}$\ and $V_{2}$\ are

\begin{equation}
V_{1}=\sqrt{f}\frac{\mid k\mid }{r^{2}}\left( \mid k\mid \sqrt{f}+\frac{r}{2}
\frac{\mathrm{d}f}{\mathrm{d}r}-f\right) ,k=j+\frac{1}{2},  \label{c14}
\end{equation}

\begin{equation}
V_{2}=\sqrt{f}\frac{\mid k\mid }{r^{2}}\left( \mid k\mid \sqrt{f}-\frac{r}{2}
\frac{\mathrm{d}f}{\mathrm{d}r}+f\right) ,k=-j-\frac{1}{2},  \label{c15}
\end{equation}%
where \(|k|=j+1/2=1,2,3,\dots\) is the total angular momentum quantum number.

Now we get that, after separation of variables, the Dirac equation reduces
to a Schrodinger-like wave equation of the form

\begin{equation}
\frac{\mathrm{d^{2}\Psi _{D}}}{\mathrm{d}r_{\ast }^{2}}+\left[ \omega
^{2}-V_{1,2}\left( r\right) \right] \Psi _{D}=0.  \label{c16}
\end{equation}%
And the effective potential $V_{1}$\ and $V_{2}$\ are mathematically
identical, that is they correspond to one equivalent potential. From the
discussion in \cite{And2}, they yield identical quasinormal modes.

\bigskip Substitute $f\left( r\right) =1-\left( \frac{r_{0}^{4}}{r^{4}}%
\right) $\ into Eq. (\ref{c14}), we get the effective potential for Dirac
field perturbation to canonical acoustic black hole

\begin{equation}
V\left( r,k\right) =\sqrt{1-\left( \frac{r_{0}^{4}}{r^{4}}\right) }\frac{%
\mid k\mid }{r^{2}}\left( \mid k\mid \sqrt{1-\left( \frac{r_{0}^{4}}{r^{4}}%
\right) }+\frac{3r_{0}^{4}}{r^{4}}-1\right) .  \label{c17}
\end{equation}

We can find that the effective potential is related to angular momentum
quantum number $\mid k\mid $.\ It vanishes both at $r\rightarrow r_{0}$\ and 
$r\rightarrow +\infty $.

\begin{figure}[h]
\centering
\includegraphics[width=0.7\linewidth]{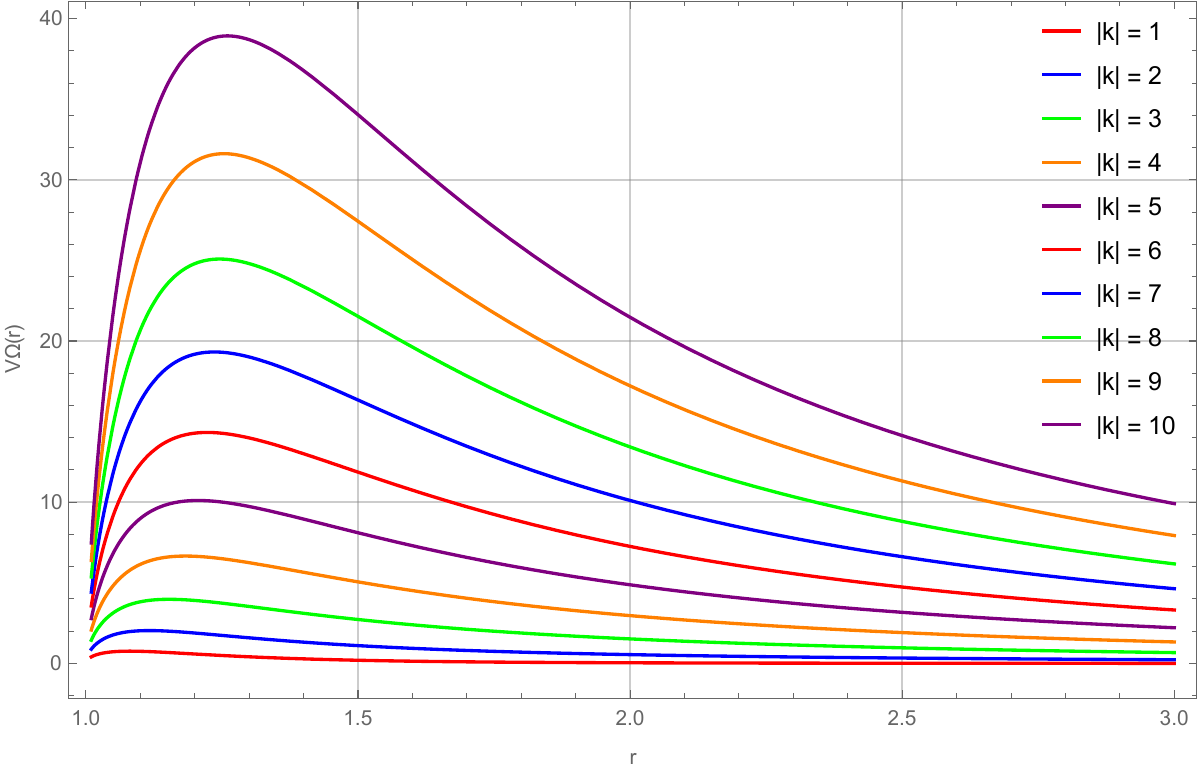}
\caption{The curve of the effective potential \(V(r)\) as a function of \(r\) when \(|k|=1\sim 10\).}
\label{fig:dirac_potential}
\end{figure}

Fig. \ref{fig:dirac_potential} shows the variation of $V\left( r\right) $\ with the coordinate $r$
for $\mid k\mid =1\sim 10$. We find that: (i) The value of effective
potentials are always positive, and they all have single peak.  (ii) The
effective potential $V\left( r,k\right) $ depends on the absolute value of $k
$, along with the increase of $\mid k\mid $, the value of the
potential increases, and the position of the peak value increases too, that is,
it moves to the right.

From Eq. (\ref{c17}), we see that when $\mid k\mid \rightarrow +\infty $,
the position of the peak

\begin{equation*}
r_{\max }\left( \mid k\mid \rightarrow +\infty \right) \rightarrow \sqrt[4]{3%
}r_{0},
\end{equation*}%
and the maximum value of the effective potential increases with $\mid k\mid $%
, as

\begin{equation*}
V\left( r_{\max },k\rightarrow +\infty \right) =\frac{2\mid k\mid ^{2}}{3%
\sqrt{3}r_{0}^{2}}.
\end{equation*}

\subsection{The QNMs of the Dirac field perturbation}

According to the effective potential in Eq. (\ref{c17}), using the sixth-order WKB approximation, we compute the quasinormal mode frequencies and analyze the stability of the canonical acoustic black hole under the perturbation of the Dirac field.

\begin{table}[h]
	\centering	
\begin{tabular}{c|c|c|c|c}
		\hline
		\(|k|\) & \(\omega (n=0)\) & \(\omega (n=1)\) & \(\omega (n=2)\) & \(\omega (n=3)\) \\ \hline		
		2 & \(1.28311-0.45682\mathrm{i}\) &  &  &  \\ \hline
		3 & \(1.80773-0.63363\mathrm{i}\) & \(1.37893-1.83396\mathrm{i}\) &  &  \\ \hline
		4 & \(2.41883-0.63674\mathrm{i}\) & \(2.06628-1.99094\mathrm{i}\) & \(1.31598-3.58222\mathrm{i}\) &  \\ \hline
		5 & \(3.05981-0.62332\mathrm{i}\) & \(2.77124-1.92423\mathrm{i}\) & \(2.17754-3.39994\mathrm{i}\) & \(1.26143-5.22826\mathrm{i}\) \\ \hline
	\end{tabular}
	
	\caption{Quasinormal modes of the Dirac field for \(|k| = 2\sim5\) computed via the sixth-order WKB approximation. Blank entries correspond to modes where the approximation breaks down (yielding positive imaginary parts) and are thus omitted.}
	\label{tb5}
\end{table}

From Table \ref{tb5}, we observe two distinct trends in the Dirac field QNMs. First, for a fixed angular momentum quantum number \(|k|\), as the overtone number \(n\) increases, the real part \(\omega_R\) decreases while the magnitude of the imaginary part \(|\omega_I|\) increases, indicating faster damping for higher overtones. Second, for a fixed overtone number \(n\) (e.g., the fundamental mode \(n=0\)), both the oscillation frequency \(\omega_R\) and the damping rate \(|\omega_I|\) increase with \(|k|\). The latter behavior is consistent with the fact that a larger \(|k|\) raises the peak of the effective potential (as shown in Fig. \ref{fig:dirac_potential}). Crucially, all imaginary parts in the reliable regime are strictly negative, which indicates that the canonical acoustic black hole is stable against Dirac field perturbations. These results are qualitatively similar to those for the Dirac field perturbation in a Schwarzschild black hole \cite{Cho}.

\vspace{0.5cm}
\noindent\textit{Remark on the choice of minimal coupling for higher-spin fields:} 
It is worth noting that non-minimal couplings between curvature and both electromagnetic and Dirac fields can arise from quantum corrections (e.g., vacuum polarization in QED) or from specific effective field theory constructions. However, in this work, we deliberately treat the electromagnetic and Dirac fields as minimally coupled test fields. This choice is motivated by two reasons: (i) it isolates the effect of the scalar field's \(\zeta R \phi^2\) coupling, allowing a clean comparison between the non-minimally coupled scalar sector and the minimally coupled baselines; and (ii) in the analogue gravity context, the electromagnetic and Dirac fields serve as probes of the effective geometry rather than as hydrodynamic degrees of freedom, making minimal coupling the natural baseline. A systematic study of non-minimal couplings for higher-spin fields in acoustic black holes will be pursued in future work.

\section{Conclusions}

In this paper, we have comprehensively investigated the stability and quasinormal modes of black hole spacetimes under perturbations from a non-minimally coupled scalar field, as well as electromagnetic and Dirac fields, using the canonical acoustic black hole as a concrete representative background. By employing the sixth-order WKB approximation, we obtained highly accurate QNM frequencies and carefully analyzed the parameter space.

Our key findings are threefold. First, using the canonical acoustic black hole as a representative case, we analytically derived a tachyonic instability threshold at \(\zeta_{\text{crit}} = -2/3\) for the monopole perturbation. More generally, the same sign analysis of the near-horizon effective potential provides a systematic route to locating the corresponding threshold in other black hole backgrounds with non-minimal scalar coupling; the presence of a negative potential region is a necessary but not sufficient condition for instability, and the time-domain evolution confirms that an exponentially growing mode does develop in the case studied here. Second, within the stable regime (\(\zeta \ge -2/3\)), we found that a positive non-minimal coupling enhances the oscillation frequency and the decay rate of the scalar perturbation, whereas a negative \(\zeta\) (approaching the critical bound) makes the perturbation decay more slowly. Third, comparing the three types of fields, we observed that the electromagnetic and Dirac fields, which are minimally coupled and gauge invariant, serve as stable baselines. The tachyonic instability appears to be a feature arising from the non-minimal curvature coupling in the scalar sector within the parameter regime explored here.

These results not only highlight the sensitivity of acoustic black hole QNMs to curvature coupling but also provide valuable insights for future investigations into analogue gravity in strong curvature regimes. Extending this framework to astrophysical black hole metrics (e.g., Schwarzschild, Reissner--Nordstr\"om) and to other analogue gravity backgrounds is a natural next step, where the background-dependent Ricci scalar \(R(r)\) will shift the numerical value of \(\zeta_{\text{crit}}\) while preserving the underlying mechanism.

\end{document}